\documentclass[aip,
 sd,%
 amsmath,amssymb,
preprint,%
]{revtex4-1}

\usepackage{graphicx}
\usepackage{dcolumn}
\usepackage{bm}

\usepackage{multirow}
\usepackage{slashbox}
\usepackage{txfonts}
\usepackage{epstopdf}

\begin{document}

\preprint{JCP, in press, doi:10.1063/1.4864657}

\title{Oxygen diffusion and reactivity at low
temperature on bare amorphous olivine-type silicate}

\author{M. Minissale}
 \email{marco.minissale@obspm.fr}
 \affiliation{LERMA-LAMAp, Universit\'e de Cergy-Pontoise, Observatoire de Paris, ENS, UPMC, UMR 8112 du CNRS, 5 Mail Gay Lussac,
95000 Cergy Pontoise Cedex, France}
\author{E. Congiu}%

\affiliation{LERMA-LAMAp, Universit\'e de Cergy-Pontoise, Observatoire de Paris, ENS, UPMC, UMR 8112 du CNRS, 5 Mail Gay Lussac,
95000 Cergy Pontoise Cedex, France}%

\author{F. Dulieu}
\affiliation{LERMA-LAMAp, Universit\'e de Cergy-Pontoise, Observatoire de Paris, ENS, UPMC, UMR 8112 du CNRS, 5 Mail Gay Lussac,
95000 Cergy Pontoise Cedex, France}

\date{\today}

\begin{abstract}

The mobility of O atoms at very low
temperatures is not generally taken into account, despite O
diffusion would add to a series of processes leading to the observed
rich molecular diversity in space. We present a study of the
mobility and reactivity of O atoms on an amorphous silicate surface.
Our results are in the form of RAIRS and temperature-programmed
desorption spectra of O$_2$ and O$_3$ produced via two pathways: O + O and O$_2$ + O, investigated in a submonolayer regime
and in the range of temperature between 6.5 and 30 K. All the
experiments show that ozone is formed efficiently on silicate at any
surface temperature between 6.5 and 30 K. The derived upper limit
for the activation barriers of O + O and O$_2$ + O reactions is
$\sim$ 150 K/k$_b$. Ozone formation at low temperatures indicates
that fast diffusion of O atoms is at play even at 6.5 K. 
Through a series of rate equations included in our model, we also address the reaction mechanisms and show that neither the Eley Rideal nor the Hot atom mechanisms alone can explain the experimental values. The rate of diffusion of O atoms, based on modeling results, is much higher than the one generally expected, and the diffusive process proceeds via the Langmuir-Hinshelwood mechanism enhanced by tunnelling. In fact, quantum effects turn out to be a key factor that cannot be neglected in our simulations.
Astrophysically, efficient O$_3$ formation on interstellar dust grains would imply the presence
of huge reservoirs of oxygen atoms. Since O$_3$ is a reservoir of elementary oxygen,
and also of OH via its hydrogenation, it could explain the observed
concomitance of CO$_2$ and H$_2$O in the ices.

\end{abstract}

\pacs{Valid PACS appear here}
\maketitle

\section{Introduction}

The chemical processes taking place in interstellar clouds can be considered the origin of
the molecular diversity in the Universe. A wealth of infrared,
millimeter- and microwave-wavelength observations has provided
evidence of rich molecular abundances within interstellar clouds.
This observational evidence for the general interstellar medium (ISM),
however, cannot be met by the known gas-phase reactions alone. That
is why surface reactions are necessarily invoked for the formation
of a growing number of molecular species. Atoms and molecules from
the gas phase accrete and gather on the cold surfaces of
interstellar dust grains, and eventually react after surface
diffusion. In fact, some of the most abundant molecules in the
Universe (such as H$_2$, H$_2$O or CO$_2$) are formed on dust
grains.\cite{Ti13} Particularly, hydrogenation of interstellar ices
is known to induce the formation of species in the solid phase and,
recently, O-atom additions as well were invoked for processes
leading to an even richer molecular diversity.\cite{Ti13} Due to the
supposed high abundance and its certain reactivity, oxygen and its
chemistry may then play a central role in astrochemistry.

Gas phase molecular oxygen has been detected\cite{Li12, La07, Go11} in
molecular clouds ($\rho$ Ophiuchi A and  OMC-1) and astrochemical
models of dark clouds predict that condensed oxygen is likely to be
a major component of apolar ices. Nevertheless, the key O$_2$
molecule remains elusive in the ISM, very probably owing to its short
lifetime,\cite{Du12} and also because it can be easily consumed at
the surface of dust grains by the two most abundant atomic species,
H and O, forming H$_2$O and O$_3$ respectively, as main products.
Water is the most abundant species in the solid phase, while solid
ozone has not yet been observed in molecular clouds. To
date, only a recent work\cite{Ch01} presents spectra compatible with
the presence of O$_3$ towards IC 5146. This apparent lack of ozone
could be explained either by a detection bias as the broad silicate
absorption band at 10 $\mu$m can easily mask the 9.6 $\mu$m ozone
feature, or by an actual lack of ozone in the ice.

The non-detection of solid ozone in dense molecular clouds is
consistent with its short lifetime on the surface of dust grains due
to its high reactivity. Mokrane et al\cite{Mo09} and Romanzin et
al\cite{Ro11} have shown that O$_3$ + H  is an efficient process
under interstellar conditions and should be able to destroy most of the
O$_3$ formed on the ice to produce water. Water formation on the
grain surfaces occurs via hydrogenation of three different oxygen
species:\cite{La13,Ti82} O, O$_2$, and O$_3$. All these routes have
been extensively studied in laboratory.\cite{Io08, Mi08, Du10, Ch12}

\begin{figure}
   \centering
  \includegraphics[width=8.6cm]{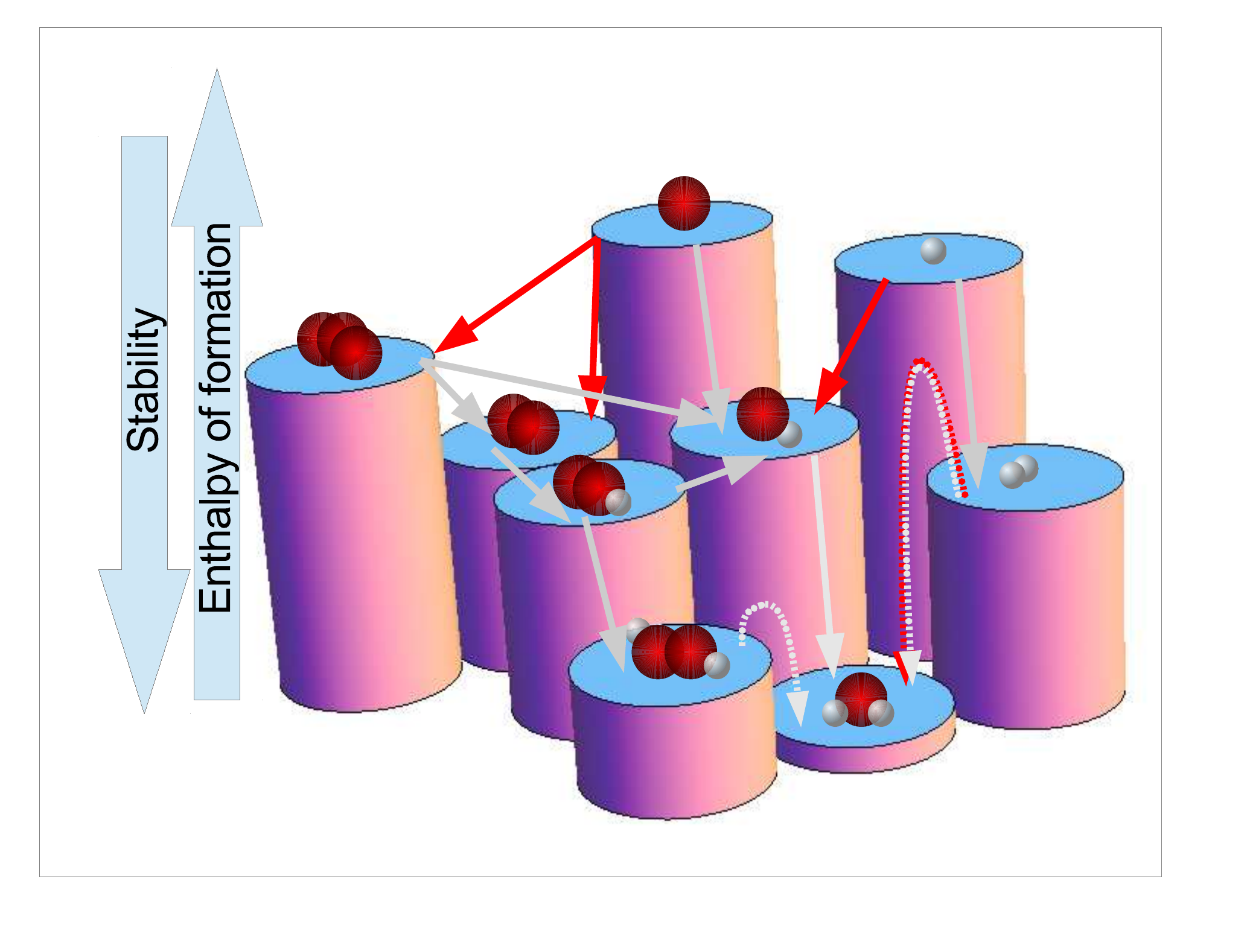}
   \caption{Schematic 3D representation of the water formation network via H-atom (grey circle) additions to O, O$_2$, and O$_3$ (red circles).
   Enthalpies of reactions grow along the vertical axis. Solid lines represent barrierless reactions, while dotted traces indicate reactions with an activation barrier.}
              \label{fig:Enthalpy}
    \end{figure}

Water is ubiquitous and its omnipresence is certainly a result of
its stability: among the species made of the 2 dominant reactive
atoms H and O, water is the most stable molecule.
Figure~\ref{fig:Enthalpy} represents all the stable molecules that
are composed of O and H atoms. Their enthalpies of formation are
represented vertically. We have drawn the different reactions
identified experimentally, of which water seems to be the
end of the chemical journey in the O and H world. However, to know
how hydrogenation of oxygen takes place for forming water, it is
necessary to understand how O$_2$ and O$_3$ are formed.

In 1930, S. Chapman discovered the mechanisms that produce the ozone layer in the Earth's stratosphere:\cite{Ch68} UV photons striking oxygen molecules (O$_2$) split them into two oxygen atoms (O); atomic oxygen then combines with O$_2$ to create ozone. In turn, O$_3$ can be dissociated by UV light into a molecule of O$_2$ and an O atom, and so on in a continuing process called the ozone-oxygen cycle, creating an ozone layer in the stratosphere.
Recently, ozone formation has been studied in the laboratory using
supra thermal oxygen atoms generated by energetic electrons or
ions.\cite{Be05,Si07,En01} Jing et al\cite{Ji12} also performed
experiments on the formation of ozone on bare silicates, but our present
work and analysis do not lead to the same conclusions. 
We attribute the detection of high temperature signals at
mass 32 a.m.u. to the decomposition of O$_3$ inside the quadruple mass
spectrometer, and not to the detection of O$_2$.
In this paper we study the surface formation of O$_2$ and O$_3$
without the addition of energy, through the reactions:\cite{Be94}
\begin{align}
O(^3P) + O(^3P) &+ M\longrightarrow  O_2 +M  \\
O(^3P) + O_2(X^3 \Sigma_{g}^-) &+ M \longrightarrow   O_3(X^1 A_1) + M
\end{align}

The substrate, made of amorphous silicate,\cite{No12} was held in
the 5~--~30~K temperature range. Sub-monolayer conditions were used
in all the experiments discussed below. This paper is organized as
follows: the experimental set-up and methods are described in the
next section. In Section 3, we present our results about O$_2$ and
O$_3$ formation. In Section 4, we present a model that simulates our results and
gives relevant energetic parameters. In the last
Section, we discuss the main conclusions and astrophysical implications of this study.

\section{Experimental methods}

Experiments were performed using the FORMOLISM set-up shown in
Figure~2, which is described elsewhere \cite{Am06, Co12}. The experiments
take place in an ultra-high vacuum chamber (base pressure 10$^{-10}$
mbar), containing a non-porous amorphous olivine-type
silicate. This sample was obtained by thermal evaporation of San Carlos olivine (Mg$_{1.8}$Fe$_{0.2}$SiO$_4$) onto a gold-coated substrate (1 cm in diameter), operating at temperatures between 6.5~K and 350~K. The surface density of adsorption sites is about the same of the one found on compact ice samples.\cite{No12} Sample preparation and surface analysis are described extensively in Djouadi et al. 2005. \cite{Dj05} The temperature of the sample T$_s$ is
computer-controlled by a calibrated silicon-diode and a thermocouple
(AuFe/Chrome K-type) clamped on the sample holder. Via a triply
differentially pumped beam, O atoms (and O$_2$ molecules) are aimed
at the cold (6.5-25K) sample. The products are probed using
temperature-programmed desorption (TPD) and Reflexion Absorption
Infrared Spectroscopy (RAIRS). The TPDs are performed by
increasing the surface temperature at 10 K/min. All mass signals are expressed in atomic mass units (a.m.u.).\\
\begin{figure*}
   \centering
   \includegraphics[width=14cm]{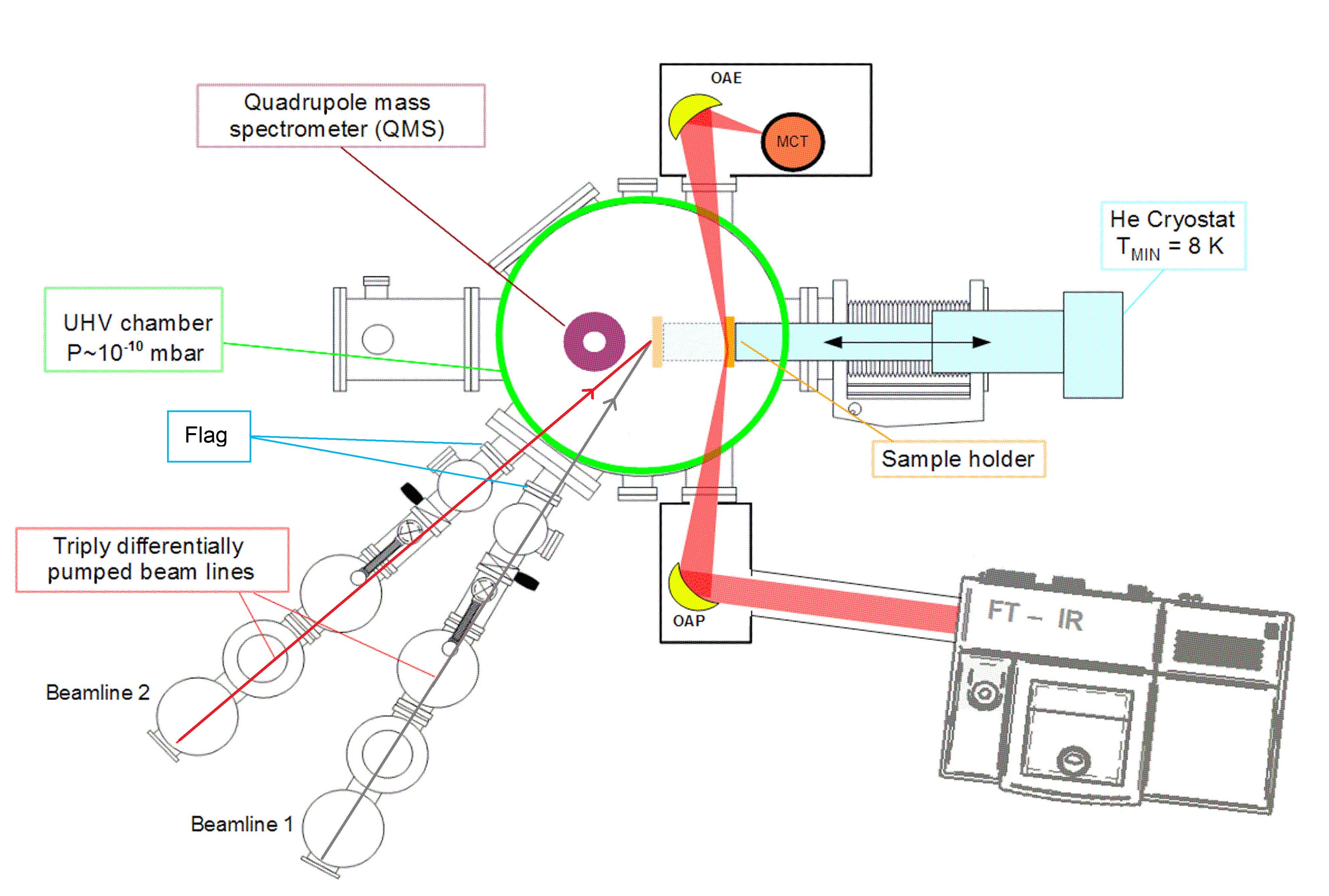}
   \caption{Schematic top-view of the FORMOLISM set-up and the FT-RAIRS facility.}
              \label{fig:setup}
    \end{figure*}
    \subsection{Oxygen beam}
Oxygen atoms are obtained by dissociating O$_2$ gas in a microwave
discharge. The dissociation fraction $\tau$ can be tuned between
45\% and 80\% by varying the microwave power. This allows us to change
the O/O$_2$ ratio sent onto the cold sample. 70\% of dissociation
(typical values used here) corresponds to a deposition with a
O$_2$/O ratio of 3/14 (14 O atoms are produced by the dissociation
of 7 molecules). It has been checked that no O$_3$ was present in
the beam. This control was carried out by two different methods. The
first control was performed by placing the quadrupole mass
spectrometer (QMS) in front of the beam and by monitoring mass-48 signal
 of the direct beam, and the one of the beam when blocked by a
metallic flag. The flag is a metal plate
used to intercept the beams before they enter the main vacuum
chamber, see Fig.\ref{fig:setup}. The signal at mass 48 was
always under the detection limit imposed by electronic noise, and
this is a first indication that no O$_3$ was present in the beam. A
second check consisted of irradiating with O+O$_2$ the surface held
at 55 K, then performing a TPD. At this temperature the residence
time of O and O$_2$ is extremely short and prevents the formation of O$_3$ through the reaction O+O$_2$, while gas phase O$_3$ sticks
and remains on the surface. A peak at mass 48 (and mass 32, see
below) in the TPD would indicate that O$_3$ was actually present in
the beam. With this second control experiment we could accurately
determine that no ozone was present in the O beam.

Also the energetic state of atoms and molecules was checked before
commencing the experiments. Molecular orbital theory predicts that
the O$_2$ molecule has two low-lying excited singlet states,
O$_2$($a^1 \Delta_g$) and O$_2$($b^1 \Sigma_{g}^+$), while the
ground state is the triplet O$_2$($X^3 \Sigma_{g}^-$) state. The
energy difference between the lowest energy of O$_2$ in the singlet
state, and the lowest energy in the triplet state is about 11340
K/k$_b$ (T$_{e}$ ($a^1 \Delta_g$ - $X^3 \Sigma_{g}^-$)= 94.3 kJ/mol,
0.98 eV).\cite{Sc03} The required energy to ionize an O$_2$ molecule
in the ground state is 12.07 eV.\cite{Ta73} This means that
electrons less energetic than this value cannot ionize
O$_2$ in the ground state, while electrons can ionize O$_2$($a^1 \Delta_g$) molecules if they are in the energy range 12.07-11.09 eV. The
same argument can be applied for atomic oxygen. O atoms in the
ground state ($^3$P) are ionized by 13.6-eV electrons,\cite{Mo93}
while in the first excited state $^1$D the minimum energy necessary for
ionization is 11.7 eV (-1.9 eV). Hence, electrons less energetic than
13.6 eV can ionize excited O atoms only.
By tuning the energy of the ionizing electrons of the
QMS, we can selectively detect ground state or excited state O$_2$ and O, as described in Congiu et al.\cite{Co09} 
Finally, we determined that the beam did not contain O or O$_2$ in an excited state, nor O$_3$ molecules.
The O beam was thus composed of at least 99\% ground-state O and O$_2$. We also recorded the
mass 16 signal in all experiments, and did not detect any signal that could be interpreted as O-atom release in the gas phase. Actually, except for the direct beam, or for a very small fraction ($<$ 2\%) due to cracking of O$_2$ and O$_3$ in the QMS, 	
we never detected any signal at mass 16. This indicates that O atoms react and
never desorb as such, but exclusively as O$_2$ and O$_3$ molecules.

\subsection{Determination of O$_2$ monolayer and flux}
\begin{figure}
   \centering
   \includegraphics[width=8.6cm]{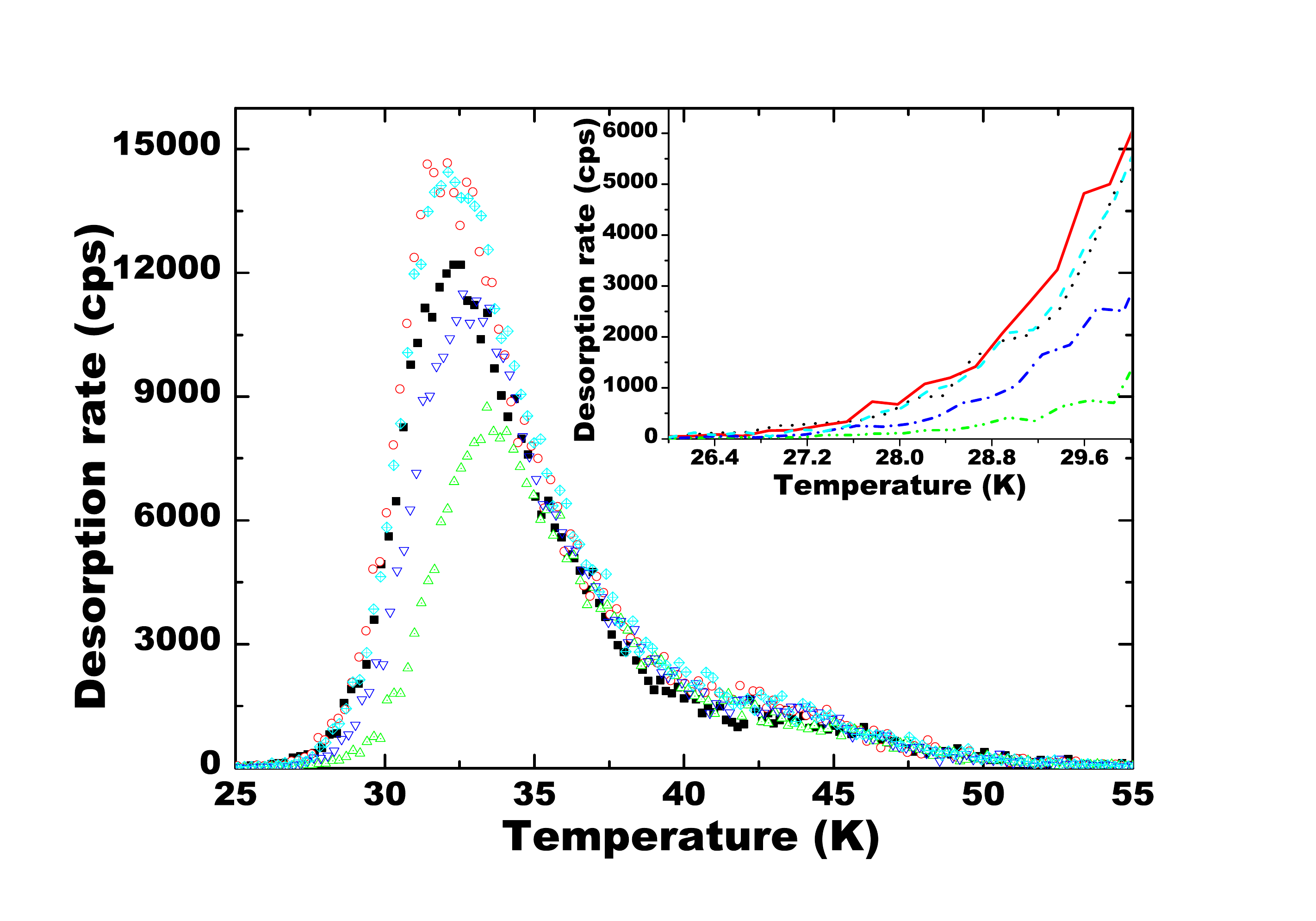}
   \caption{TPD mass spectra at mass 32 after 240, 270, 300, 330, and 360 s of O$_2$ exposure on silicate held at 10 K. The inset shows a magnified view of the leading edges between 26 and 30 K.}
              \label{fig:O2cal}
    \end{figure}

The technique used to determine the O$_2$ flux was adapted from
Kimmel et al 2001.\cite{Ki01} The O$_2$ flux was calibrated by
saturation of the first O$_2$ monolayer\cite{Am07,No12} as shown in
Fig.~\ref{fig:O2cal}. The method consists of depositing
different amounts of O$_2$ -- under identical conditions of flux --
on the surface maintained at the same temperature (in this case T$_s$= 10~K). With the increase in the doses deposited on the surface, the
TPD curves gradually broaden towards lower temperatures. In fact, as the surface coverage increases,
 the molecules are adsorbed in less
tightly bound adsorption sites, namely the desorption temperature
T$_{des} \propto$ desorption energy E$_{des}$ (with peaks growing in
height too). When the leading edge of the TPD curves (the left side
of the curves shown in the inset of Fig.~\ref{fig:O2cal})
stops shifting towards lower temperatures, it means that all
the adsorption sites on the surface are occupied, and any other
incoming molecule is adsorbed on top of the first layer of molecules already
adsorbed on the surface. This is when TPDs exhibit a 0$^{th}$ order
desorption, the maxima of the desorption peaks increase and start
shifting towards higher temperatures with increasing doses. The unit
of coverage we adopt -- the monolayer (ML) -- is defined as a single layer of atoms or molecules adsorbed on a surface. 
Here the monolayer
coverage of O$_2$ molecules adsorbed on silicate occurs after
315$\pm$30 seconds of O-beam exposure. The integrated signal
(expressed in counts per second) over the temperature of the TPD
curve after a 315-second exposure corresponds to 1 ML. The same dose
is necessary to fill the monolayer if the substrate is made of compact
amorphous water, which means that the monolayer is achieved with the
same number of molecules (i.e., a similar wetting of the two
surfaces occurs). This is not the case (about half of the above
dose is needed) if a substrate of graphite is used. Due to the similarity between the 
behaviours of the silicate and water ice substrates (which implies
that the density of adsorption sites is of the same order of
magnitude), we can estimate that 1 ML on the silicate sample is
10$^{15}$ molecules cm$^{-2}$, within the uncertainty of this
technique (about 15\%). In this work, the doses are expressed in
terms of O$_2$ units, which means that 1 ML may also represent 2
layers of O atoms or 0.66 layers of pure ozone.
       \subsection{O$_3$ detection efficiency and calibration}
\begin{figure*}
   \centering
   \includegraphics[width=14cm]{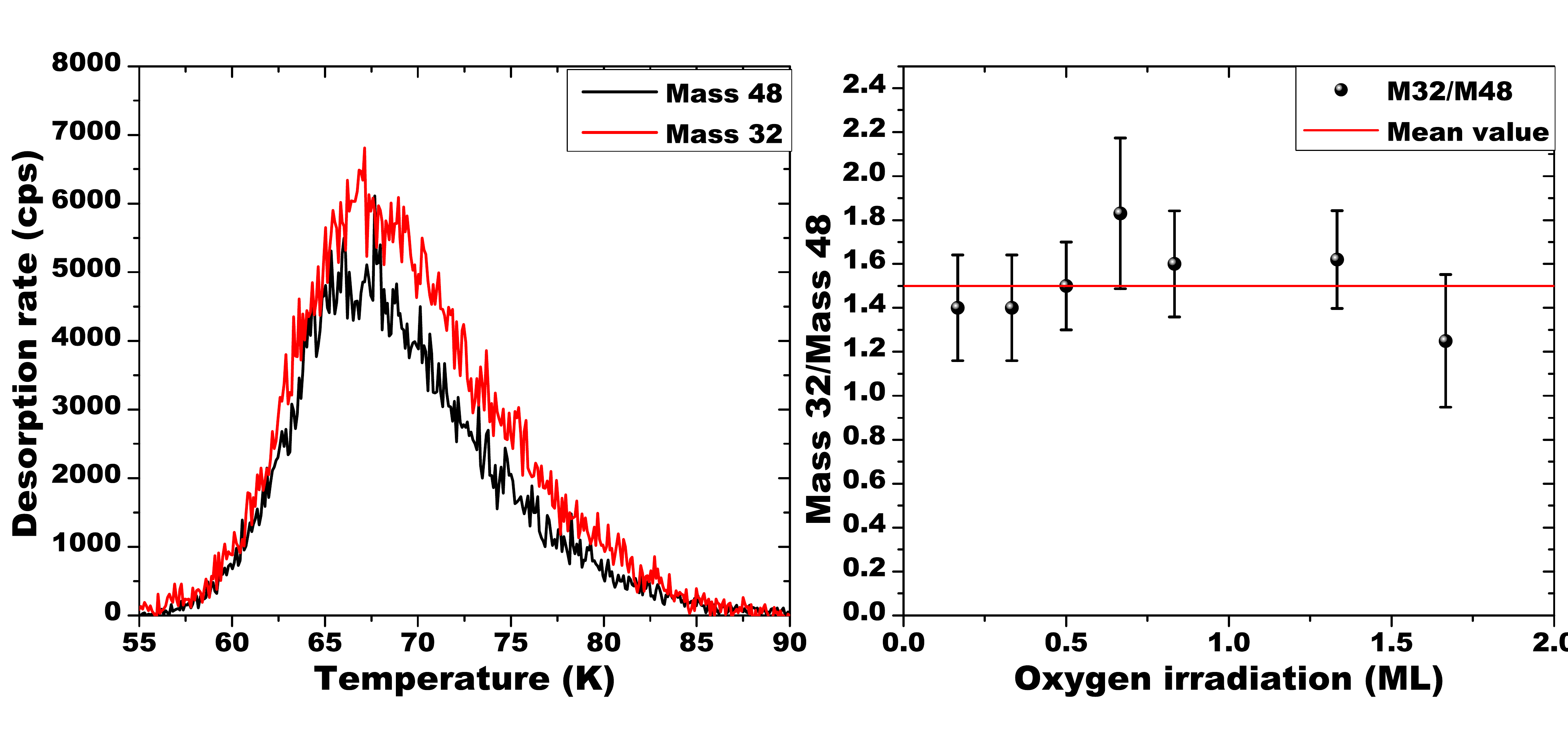}
   \caption{\textit{Left panel:} Ozone TPD curves at mass 32 and 48 between 55 and 90 K after deposition of 5 minutes of oxygen atoms on silicate held at 10 K.
   \textit{Right panel:} Ratio of the integrated area of mass 32 and mass 48 TPD peaks (55-90 K) as a function of different doses of oxygen atoms.}
              \label{fig:M32_M48}
    \end{figure*}

When using a beam of O$_2$, it is easy to calibrate the flux and
understand when the saturation of one ML occurs. This is not the
case for a deposition of O$_3$. To calibrate one ML of O$_3$, it is
necessary to study and evaluate the cracking pattern and the
detection efficiency of ozone by the QMS, with respect to the
well known O$_2$ detection. As already seen in Mokrane et al
2009,\cite{Mo09} the ozone signal is simultaneously detected at mass
48 and at mass 32 (O$^{+}_{2}$ fragments). This is because the
dissociation of O$_3$ is energetically more favorable than its
ionisation  (3.77 eV vs 12.53 eV). When O$_3$ enters the QMS head,
it can undergo different processes:
\begin{itemize}
\item  O$_3$ (+ e$^{-}$) $\longrightarrow$  O$^{+}_{3}$  ($\Delta$H$_f$ = 12.97 eV)
\item O$_3$ (+ e$^{-}$) $\longrightarrow $ O$^{+}_{2}$ + O ($\Delta$H$_f$ = 13.17 eV)
\item O$_3$ (+ e$^{-}$)  $\longrightarrow $ O$^{+}$ + O$_2$ ($\Delta$H$_f$ = 14.72 eV).
\end{itemize}
The left panel of Fig.~\ref{fig:M32_M48} shows the TPD spectra at mass 32 and
48 between 55 K and 90 K after a deposition of 5 minutes of oxygen
atoms on silicate. The two traces exhibit the same shape, namely,
the mass32/mass48 ratio remains constant
(right panel of Fig.~\ref{fig:M32_M48}). The two curves are clearly due
to the desorption of the same parent molecule (ozone) formed on the
substrate. The assignment of the high-temperature peak at mass 32 is
the main experimental difference between the interpretation of the
data in this work and in Jing et al 2012's.\cite{Ji12} They
attribute the O$_2$ peak desorption between 55 and 90 K to O
recombination and subsequence desorption of molecular oxygen. In the
present study, we determined that the mass-32 peak is due to ozone
desorption and its fragmentation upon detection. The deposition of
ozone from ex-situ synthesis confirms this fact: the amount of
desorbing ozone can be monitored either via mass 32 or mass 48. By
computing the ratio between the two signals (mass32/mass48) after
deposition of different doses, as shown in
Fig.~\ref{fig:M32_M48} (right panel), a mean value of 1.5 was found. This
fact led us to monitor ozone by the signal at mass 32, instead
of that at mass 48, to have a better signal-to-noise ratio.

Due to its important dissociative ionization, ozone detection efficiency
 has to be determined for every single QMS.
Actually, the results that gave the idea of the present work were
obtained during two periods of experiments, when two different QMSs were used. Even thought the settings of the two instruments were, at any
time, exactly the same, the O$_3$/O$_2$ detection efficiencies found
were different up to a factor of 30\%. For the sake of consistency,
however, all the experimental values presented here were obtained
with a constant O$_3$/O$_2$ detection efficiency. It should also be
noted that, for this specific molecule (O$_3$), any other correction
factor among those present in the literature would have be wrong in
our case.
To estimate the O$_3$/O$_2$ detection efficiency at mass 32, it is
necessary to compare the area under the TPD curve of one ML of O$_2$
and that of one monolayer of O$_3$. Therefore, we also needed to
determine for what O-exposure time we reached a complete monolayer of
ozone.
   \begin{figure}
   \centering
   \includegraphics[width=8.4cm]{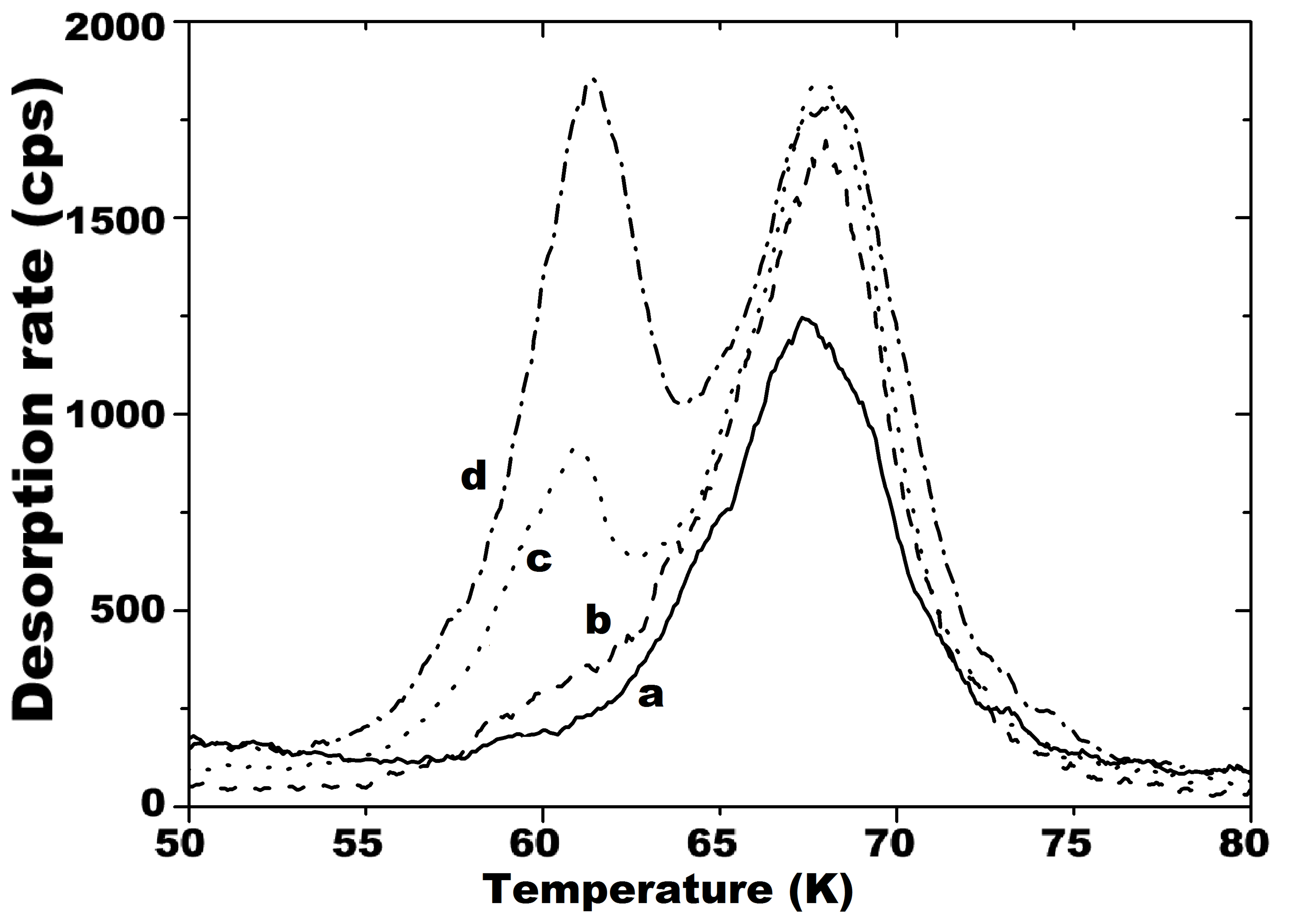}
   \caption{Ozone TPD curves between 50 and 90 K after four depositions of oxygen atoms of (from \textit{a} to \textit{d}) 6, 8, 10, and 12 minutes on silicate held at 10 K. Curve \textit{b} represents desorption of a complete ozone monolayer, while \textit{c} and \textit{d} exhibit also another peak at $\sim$60 K due to desorption of O$_3$ from the second layer.}
              \label{fig:ozonecal}
    \end{figure}
    
To calibrate the ozone monolayer, we adopted the same first
layer-saturation method used for O$_2$. To do so, we gradually increased 
the amount of ozone formed on the surface (via O+O$_2$
reaction), until the second-layer desorption peak appeared. In fact,
the second-layer desorption peak is a clear signature that the first
monolayer has been completed and that a new layer is being grown. In
Fig.~\ref{fig:ozonecal}, we show four TPD curves of ozone
obtained by increasing the exposure time of oxygen atoms (from
curve \textit{a} to \textit{d}). The saturation of the first layer
corresponds to trace $b$, and the appearance of the second peak is
observed in trace $c$. The apparent inconsistency between
the TPD peak intensity in Fig.~\ref{fig:M32_M48} (left panel) and
Fig.~\ref{fig:ozonecal} is due to different QMS settings.
Nevertheless, we cannot be certain that the adsorption site
density for ozone is the same seen for molecular oxygen. Also, by
applying the derived efficiency detection factor ($ef$), we found
that the number of O atoms desorbed as O$_2$ and O$_3$ were
fewer than the number of O atoms exposed. This may seem non-consistent
with the results found in a previous study on water ice.\cite{Mi13}
We demonstrated, however, that the missing atoms in the case of the
silicate substrate, are due to the prompt release of molecules upon
formation, the so-called chemical desorption.\cite{Du13} Therefore,
the amorphous silicate substrate is not suitable for calibration
measurements.

When working on a water ice substrate instead, the
surface-saturation method gives a reliable detection efficiency
and an exact linear relation between the products and the deposited
species. For this reason, the water ice substrate assures that the
efficiency factor $ef$ is correctly estimated. Moreover, the density
of adsorption sites for O$_2$ and O$_3$ is identical. We then expect
that the total number of O atoms and O$_2$ molecules sent onto the
water ice substrate is conserved and it is equal to
O+O$_2$+$ef\times$O$_3$. To check the conservation of O atoms, all
the species are normalized to the O$_2$ signal (the O signal is
divided by 2 and that of O$_3$ is multiplied by 1.5). Different
doses of O+O$_2$ are then sent onto the icy substrate and a typical
set of temperature-programmed experiments is performed.

\begin{figure}
   \centering
   \includegraphics[width=8.6cm]{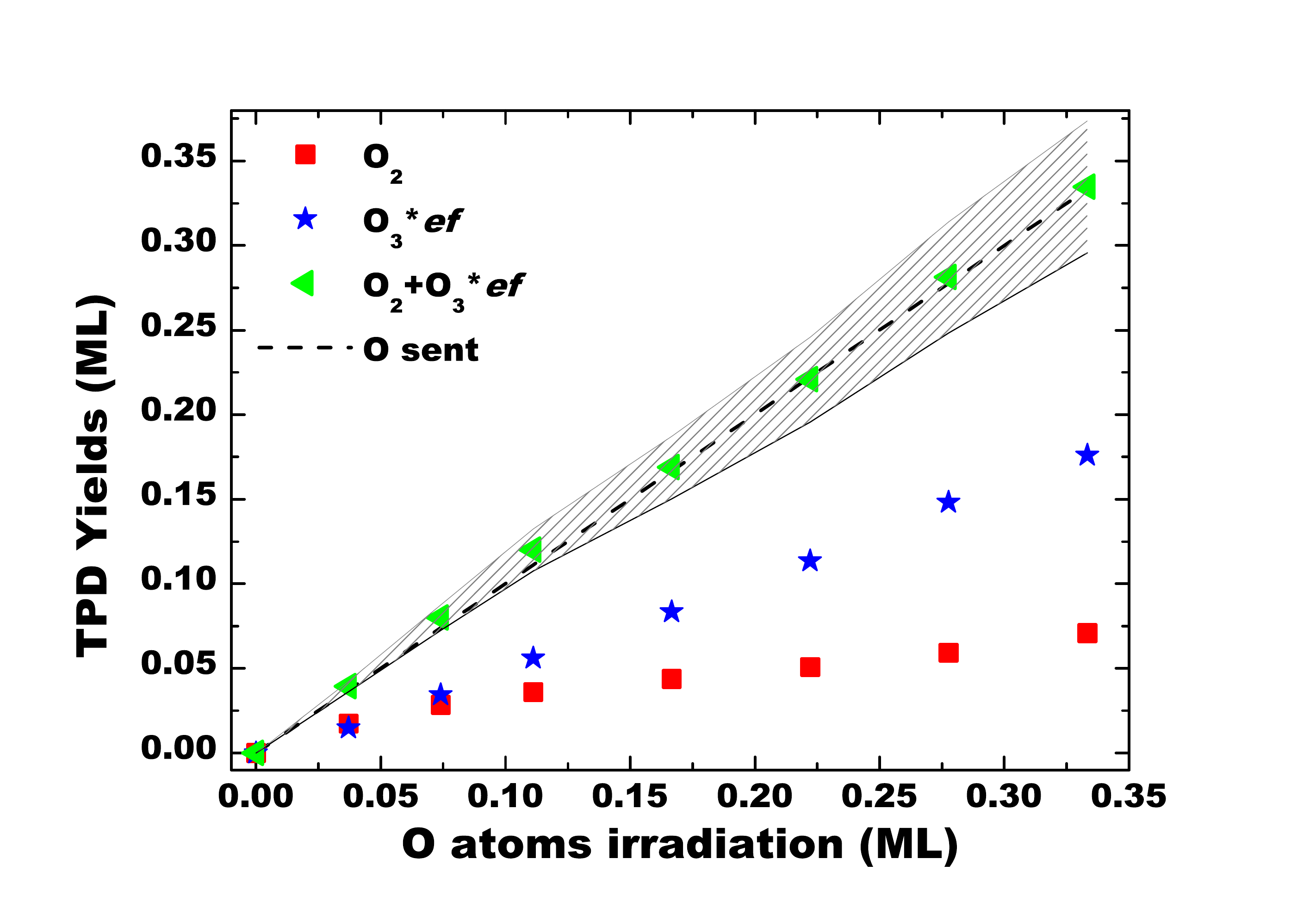}
   \caption{Integrated areas (in ML) of the TPD peaks of O$_2$ (red squares) and O$_3$*\textit{ef} (blue stars) vs O-atom fluence. Green triangles: sum of O$_2$ and O$_3$*\textit{ef}. The dashed line represents the total amount of oxygen atoms deposited on the surface. The shaded region indicates a $\pm$ 30\% variation of the efficiency factor \textit{ef}.}
              \label{fig:eff_fac}
    \end{figure}

The dashed line in Fig.~\ref{fig:eff_fac} corresponds to TPD yields after pure
O$_2$ deposition carried out with the undissociated beam. The red
squares in Fig.~\ref{fig:eff_fac} were obtained through
integration of the area under the O$_2$ TPD curves of mass 32
between 25-50 K. The blue stars correspond to the integration of the
ozone signal multiplied by the efficiency factor. The green
triangles are obtained by adding the O$_2$ and the corrected O$_3$
($ef\times$O$_3$) contributions. These contributions lie -- within the
experimental errors -- on the line given by the total amount of
deposited O atoms, which indicates that both the efficiency factor
used and the monolayer calibration are reliable. It must also be
noted that the O$_3$/O$_2$ ratio varies greatly with the coverage,
so if the efficiency factor is not correctly estimated, the total
O$_2$ + O$_3$ yield cannot be proportional to the initial dose. To
show the reliability of our $ef$ factor, in
Fig.~\ref{fig:eff_fac} we drew a shaded area indicating a
$\pm30\%$ variation of $ef$. Finally, the two independent
estimations of the monolayer coverage and the fact that the number
of O atoms is conserved, regardless of the ratio of O and O$_2$ sent
onto the surface (and the O$_2$ and O$_3$ ratio desorbing from the
surface), clearly suggests that our calibrations were correct.

\section{Experimental Results}

\begin{figure}
   \centering
   \includegraphics[width=8.6cm]{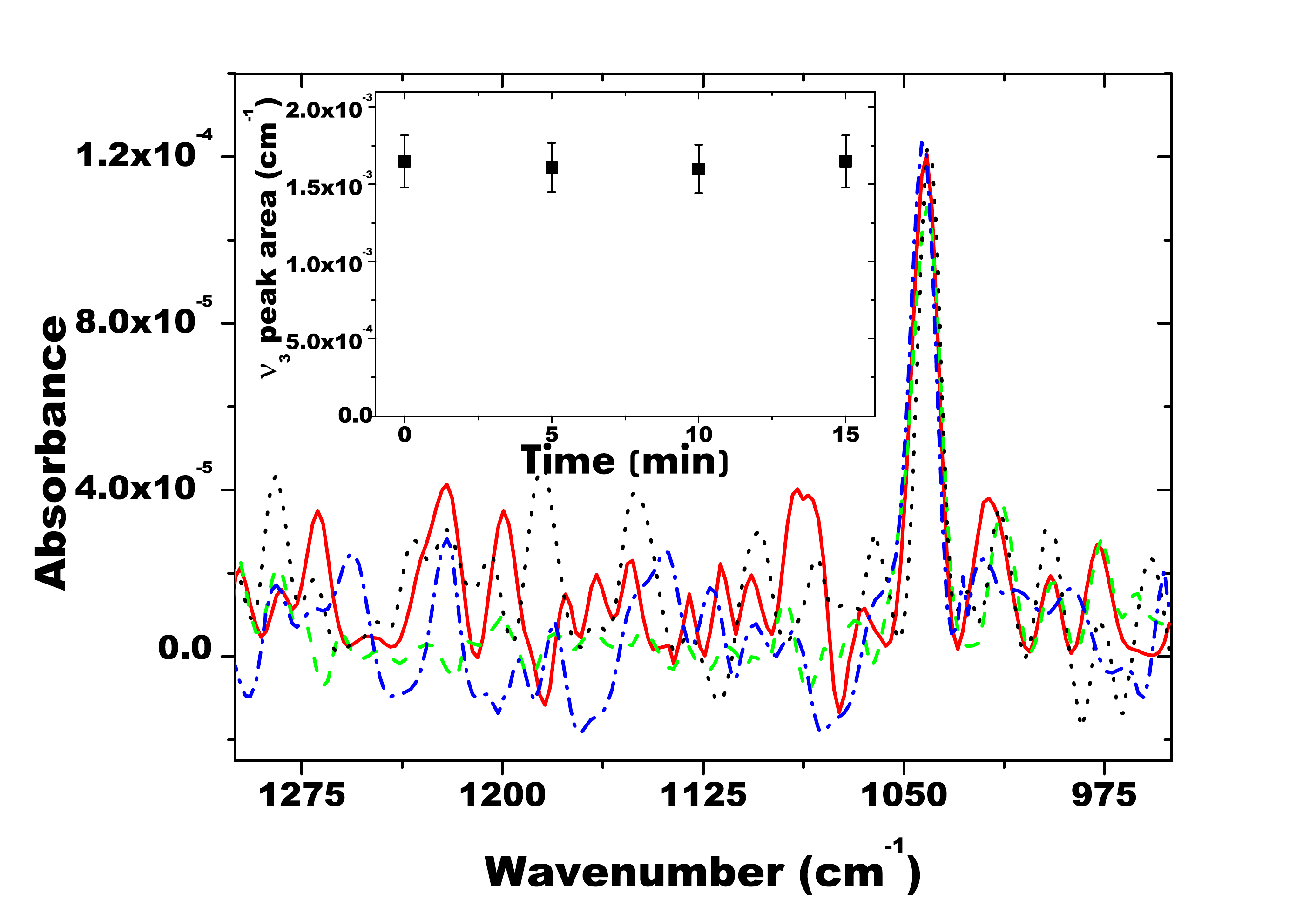}
   \caption{Series of RAIR spectra recorded at different times after deposition of 0.3 ML of O atoms at 6.5 K; the absorption bands at 1043 cm$^{-1}$ is because of the $\nu_3$ asymmetric stretching mode of O$_3$. Inset: $\nu_3$ band integrated area as a function of wait time (0, 5, 10, and 15 min).}
              \label{fig:Rairs_Time}
    \end{figure}

\begin{figure*}
   \centering
   \includegraphics[width=14cm]{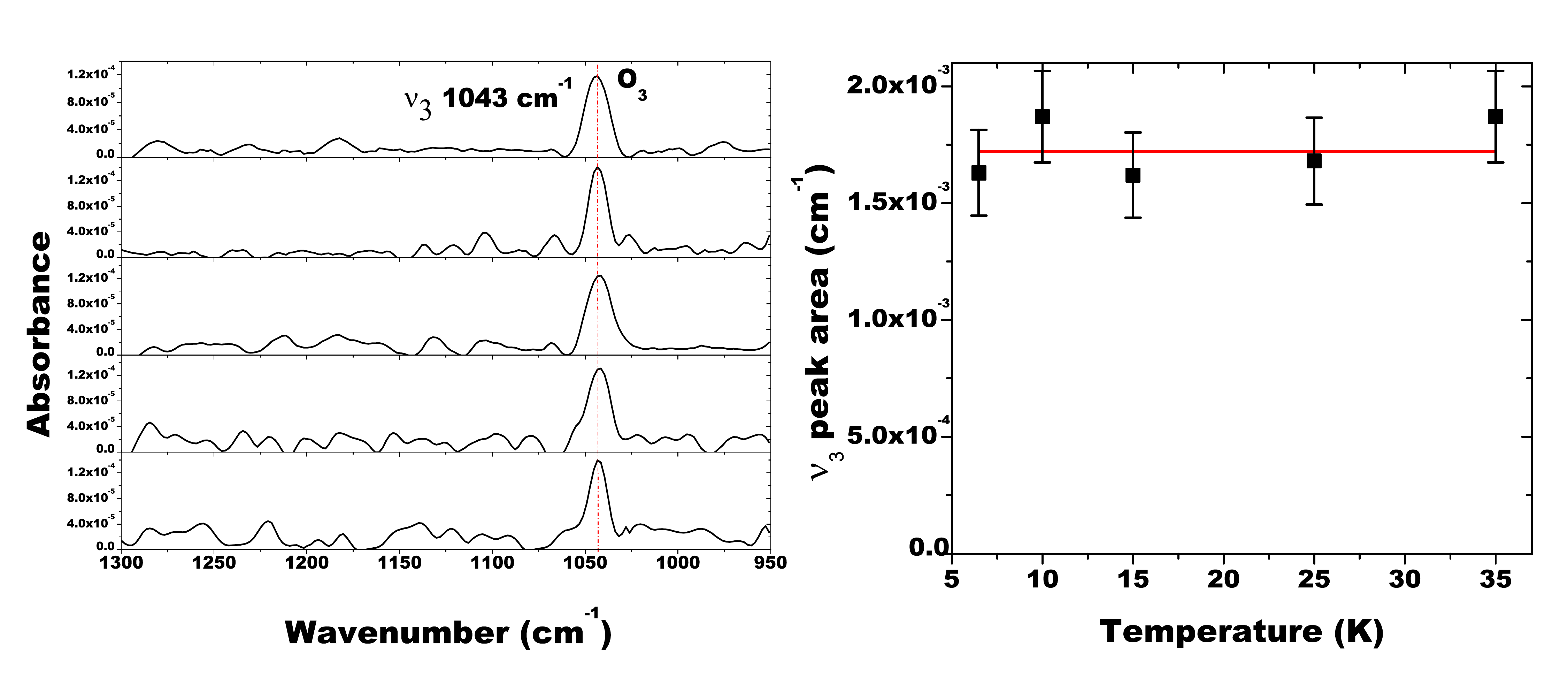}
   \caption{\textit{Left panel:} RAIR spectrum obtained at different T$_s$ (bottom to top, 6.5, 10, 15, 25 and 35 K) after deposition of 0.3 ML of O atoms at 6.5 K; the absorption band at 1043 cm$^{-1}$ is due to the $\nu_3$ asymmetric stretching mode of O$_3$. \textit{Right panel:} integrated area of the ozone band as a function of surface temperature. The red solid line represents the mean value of the five integrated band areas.}
              \label{fig:Rairs_Temp}
    \end{figure*}

All the experiments described below indicate that ozone is formed
efficiently on silicate at any deposition temperature of the surface between 6 and 25 K.
The first evidence of ozone formation is in the infrared
spectrum recorded after depositing 0.3 ML of atomic oxygen on silicate held
at 6.5 K (Fig.~\ref{fig:Rairs_Time}).
0.3 ML also represents the lower detection limit of the ozone band
in our IR spectrometer. Working with low coverages is the key to
understanding what mechanism is at play in ozone formation. There
are mainly two mechanisms that may lead to ozone. The Eley-Rideal
mechanism (ER) occurs when one of the molecules already adsorbed
promptly reacts with a particle coming from the gas phase, before being adsorbed on the surface. The
Langmuir-Hinshelwood mechanism (LH) describes the formation of
molecules on a surface when two adsorbed reaction partners react because of 
the diffusion of at least one of them. ER is independent of
T$_s$, and it becomes more efficient with the increase in
surface coverage. At high coverages (more than one ML) it becomes
the most probable mechanism. Conversely, the LH mechanism is
initiated by the mobility of the species and is very sensitive to
T$_s$. Whenever the diffusion is fast enough, it may be efficient
also at low coverages (see below).
In Fig.~\ref{fig:Rairs_Time}, we show the $\nu_3$ asymmetric
stretching mode of $^{16}$O$_3$ at  1043 cm$^{-1}$.\cite{Cl66, Bu94} The weak bands at 1103 and 700.9 cm$^{-1}$
are not visible due to our experimental conditions. The presence of
the $\nu_3$ band indicates that ozone was formed already at 6.5 K.
Moreover, this band does not evolve with time (simply by waiting 5,
10, or 15 minutes at 6.5 K), as shown in the inset of
Fig.~\ref{fig:Rairs_Time}, where the squares represent the
integrated area under the peaks. In the left panel of
Fig.~\ref{fig:Rairs_Temp} we show how the ozone band evolves
with temperature. The squares in Fig.~\ref{fig:Rairs_Temp} (right panel)
represent the total absorbance by integration of the O$_3$ band as a
function of surface temperature. The intensity of the band does not
change within the limits of the error bars. This indicates that no
 ozone was formed during the heating, when the diffusion and
reactivity of the ad-atoms (if present) should increase.  In fact,
the reactions leading to ozone
formation had already occurred at the deposition temperature via the LH mechanism.
However, due to the size of the error bars, a small increase of the
ozone band could have still been possible. We estimated that an
upper limit for the fraction of extra ozone formed during the
heating is 15\%, a value that we will use below in the discussion.

\begin{figure}
   \centering
   \includegraphics[width=8.6cm]{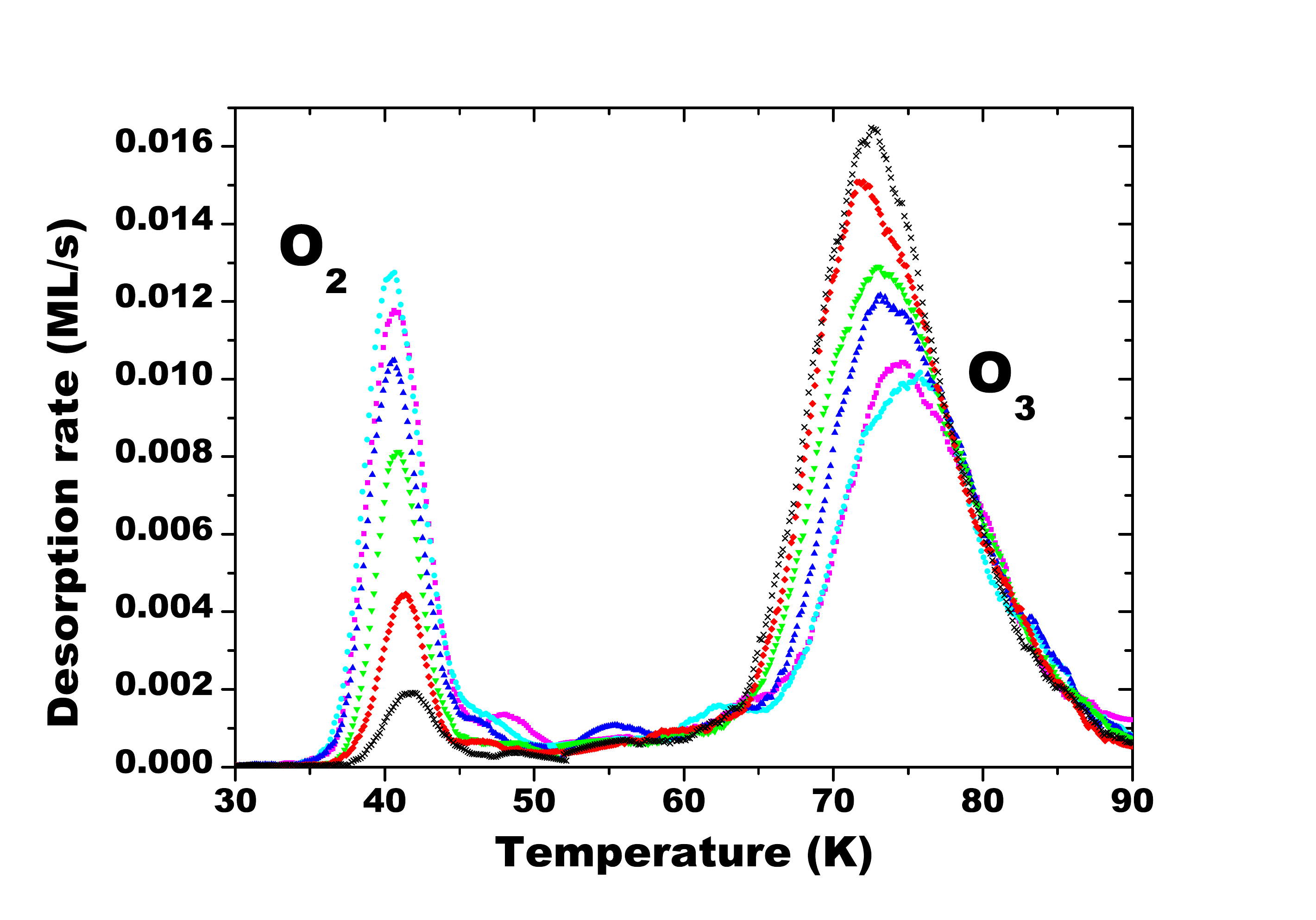}
   \caption{Series of 6 TPD curves after deposition of 0.3 ML of O atoms on silicate held 8, 10, 15, 20, 25, and 30 K. The low temperature peak (O$_2$) decreases with deposition temperature while the high temperature peak (O$_3$) grows bigger with increasing deposition temperatures.}
              \label{fig:TPDtemp}
    \end{figure}

\begin{figure}
   \centering
   \includegraphics[width=8.6cm]{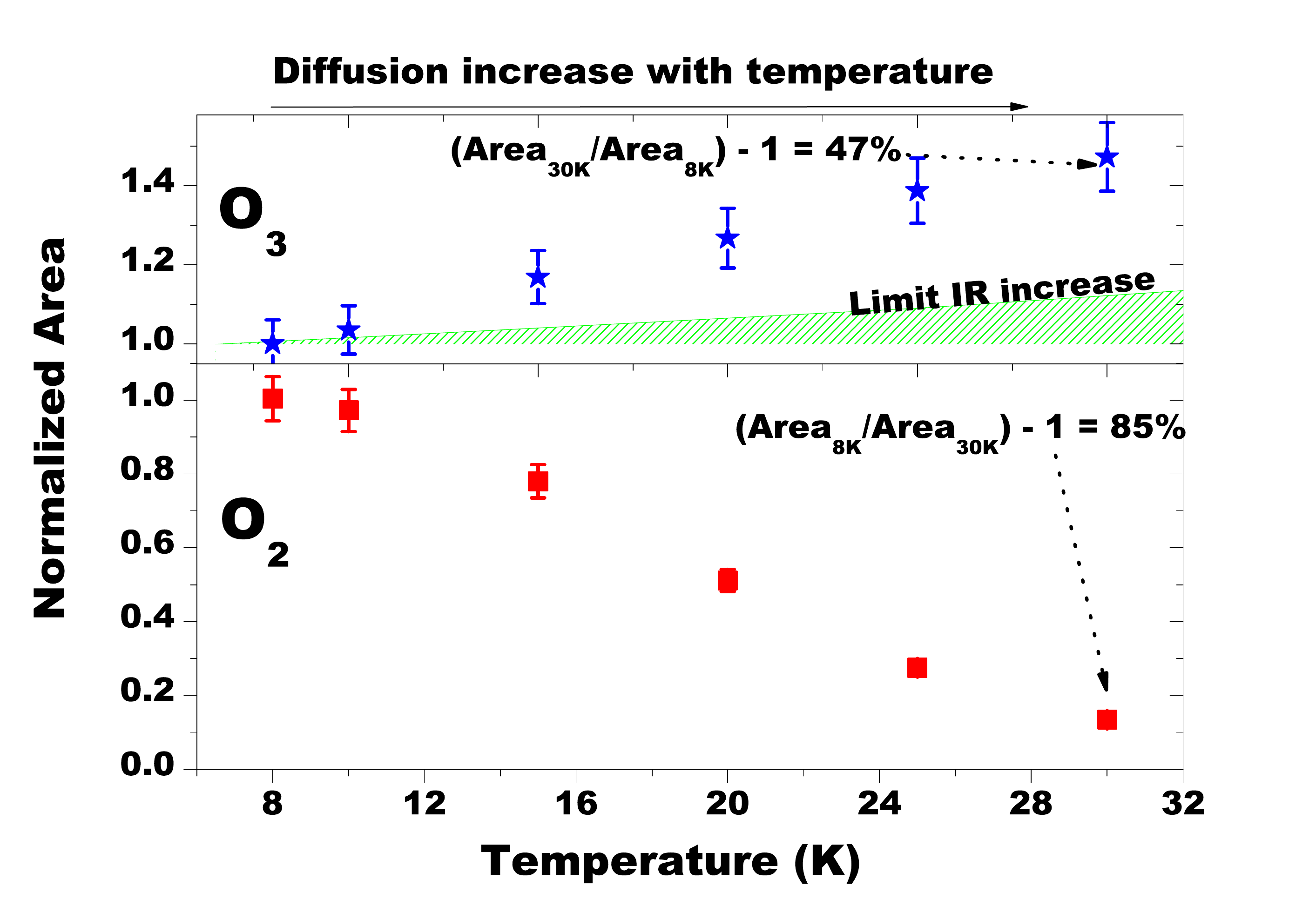}
   \caption{Integrated peak areas of O$_2$ (lower panel, red squares) and O$_3$ (top panel: blue stars) yields (obtained in the experiments shown in Fig.~\ref{fig:TPDtemp}) vs deposition temperature. The peak areas were normalized w.r.t. the TPD yields after deposition at 8 K. Green pinstriped region: range of values of the O$_3$ signal increase due to thermal diffusion derived from IR spectra.}
              \label{fig:RairsTPD}
    \end{figure}

To disentangle the ER from the LH mechanism, one
should vary the coverage, since ER is very sensitive to it, and
the temperature of the surface since the LH mechanism efficiency is
governed by the diffusion at a given temperature.  We then deposited
equal amounts of O+O$_2$ for a total of 0.3 ML at different T$_s$,
and performed a TPD at 10 K/min after each deposition. The resulting
TPD traces are presented in Fig.~\ref{fig:TPDtemp}. In each
mass spectrum, two desorption peaks appear: O$_2$ desorbs between 35
K and 50 K, while ozone desorption is observed between 55 K and 75 K
(directly at mass 48, or via the O$_2$$^+$ fragments at mass 32). O
desorption was never observed. The height of the peaks (proportional
to the amount of the species formed on the surface) changes
depending on the coverage and on the surface temperature.
Fig.~\ref{fig:TPDtemp} summarizes the outcome of six TPDs, in
which the coverage was fixed (0.29 $\pm$ 0.03 ML) and the deposition
temperature varied between 8 K and 30 K. We can observe, from curve
to curve, a clear change in the O$_3$/O$_2$ ratio. This effect is
due to the temperature of the silicate substrate and is a sign of
the role of diffusion in the formation of ozone. In fact, with
increasing surface temperature, the mobility of O atoms is favored,
ozone formation is more efficient, and the O$_3$/O$_2$ ratio
increases. Each new adsorbed atom -- if the diffusion is fast -- is
able to scan the surface to react with O$_2$ to form O$_3$, or with
another absorbed O atom to form O$_2$ (that, in turn, will also be
transformed into O$_3$ by the next incoming and mobile atom). In
this scenario, almost all O atoms and O$_2$ molecules are
transformed into O$_3$ molecules. On the contrary -- if the
diffusion is slow -- an oxygen atom has not enough time to scan the
surface and react with an adsorbed O$_2$. Another O atom then comes
and more O$_2$ is formed via the O+O reaction. A reduced
mobility leads to the accumulation of O atoms on the surface, the
probability for an O atom to meet another O atom raises, and
eventually the O$_2$ formation is favored. \\
By comparing RAIRS and TPD results, it is possible to see how the
diffusion of O atoms changes the O$_3$/O$_2$ ratio. As stated above,
we have assumed that an increase of 15\% of the ozone yield may have
occurred during the heating from 6.5 to 35 K (see
Fig.~\ref{fig:Rairs_Temp}). From TPD results, however, we
obtain a variation of 47\% between the ozone yields after O
deposition performed at 8 K and the one performed at 30 K. This
indicates that, taking into account the possible 15\% contribution
due to the heating, there is a 32\% (47 from TPD, -15 from RAIRS)
difference between TPD experiments carried out at T$_{s}$=8 and 30
K. In the upper panel of Fig.~\ref{fig:RairsTPD} we show
TPD (blue stars) and RAIRS (green shaded region) normalized peak
areas (yields) of ozone. TPD variations (also considering the error bars) are
greater than the ones in RAIRS data. This difference is clearly an
effect due to T$_{s}$, i.e., variations of O atom mobility on the
silicate surface. In the lower panel of
Fig.~\ref{fig:RairsTPD} we show the O$_2$ yield variations in
TPD experiments after O exposures at various T$_{s}$, normalized
with respect to the TPD yield after deposition of O atoms at 8 K.
In depositions carried out at 30 K, only 15\% of the amount of O$_2$ formed
at 8 K was observed.

\begin{figure}
   \centering
   \includegraphics[width=8.6cm]{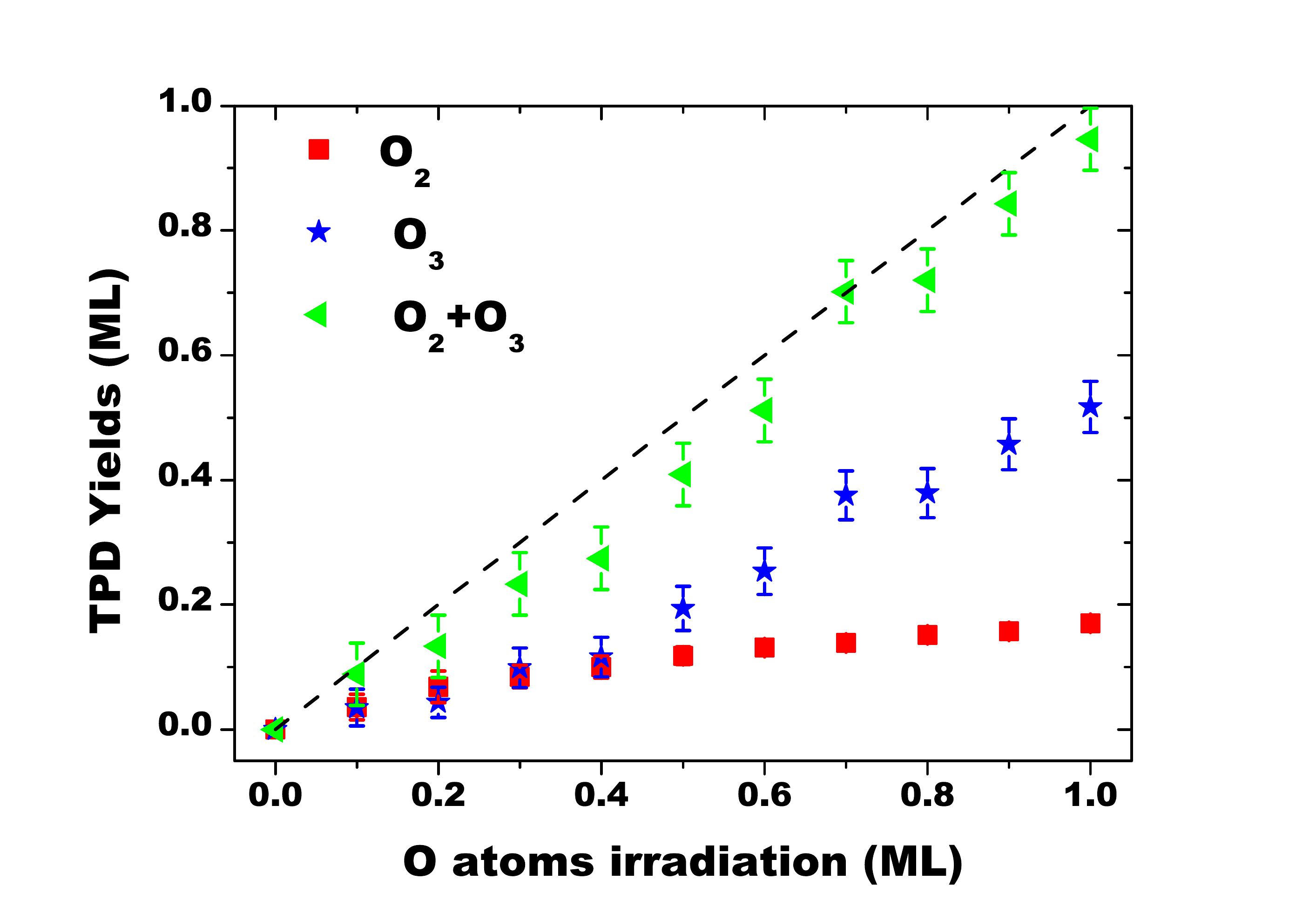}
   \caption{Integrated peak areas of O$_2$ (red squares) and O$_3$ (blue stars) yields  as a function of O-atom dose at 10 K. Green triangles: sum of O$_2$ and O$_3$ yields. The dashed line represents the total amount of oxygen atoms deposited on the surface.}
              \label{fig:O3cov}
    \end{figure}

Fig.~\ref{fig:O3cov} shows the integrated peak areas of O$_2$ and
O$_3$ TPDs as a function of the amount of deposited O atoms.
Exposures were performed at T$_{s}$=10 K with coverages going from
0.1 to 1.0 ML. Red squares represent the molecular oxygen yield and
blue stars represent the ozone yield. The O$_2$ production reaches a 
value of little less than 0.2 ML, with a growing rate diminishing
with the coverage. On the other hand, the ozone yield increases
with the coverage and reaches a value of about 0.5 ML. Green triangles in
Fig.~\ref{fig:O3cov} represent the sum of the ozone and
oxygen integrated peak areas while the dashed line is the total
amount of oxygen atoms sent onto the surface. The discrepancy
between the total yield of products (O$_2$ + O$_3$) and the dashed line
-- indicating a non-conservation of oxygen atoms -- is due to the
the chemical desorption of oxygen molecules.\cite{Du13} It is clear,
from Fig.~\ref{fig:O3cov}, that the difference between the
number of atoms sent onto the surface and those detected is maximum in the range of
coverages between 0.2 and 0.5 ML, i.e., in the low coverage regime
where chemical desorption is more effective.

To have a better understanding of the mechanisms occurring
on the silicate surface, we have developed a model that we present
in the next section. Our model was conceived to fulfill the
following experimental evidences:

\begin{enumerate}
 \item The O$_3$/O$_2$ ratio depends both on the coverage and on the surface
 temperature.
 \item At T$_s$ = 6.5 K, with 0.3 ML of O-atom coverage, more than 85\% of ozone is formed during the deposition
 phase.
 \item Experimental data confirm that chemical desorption of O$_2$ molecules occurs, and its efficiency seems to decrease with coverage. To simplify our model, however, we have assumed a constant chemical desorption rate.
 \end{enumerate}

\section{Model}

The O$_2$ and O$_3$ formation can occur via the following exothermic
reactions:

\begin{align}
O + O \longrightarrow  O_2 \\
O + O_2 \longrightarrow  O_3\\
O + O_3 \longrightarrow  2 O_2
\end{align}
The exothermicity of these reaction is 5.2 eV, 1.1 eV and 4.0 eV,
respectively. If reaction (5) were efficient, it would cause a decrease in the ozone
amount and double an increase in molecular oxygen. However, taking
into account the fact that the ozone production efficiency is close
to unity at high temperature or high coverage, the third reaction is
apparently not efficient under our experimental conditions. Hence we
can neglect it and assume that this is probably due to a barrier to
the O + O$_3$ reaction. We will include this reaction later in the discussion to
estimate the height of the barrier, and check whether our initial
assumption is reasonable.

By supposing that the reactions leading to O$_2$ and O$_3$ are
governed by the density of species on the surface and by T$_s$, we
can model the processes through a series of rate equations. We
tried to fit our data by using different approaches, and tested
different hypothesis. Our model includes both ER and LH mechanisms.
It also allows reactions to occur during the heating ramp, as well
as during the exposure phase, even if we know -- thanks to the IR
spectra -- that this contribution should be small. In addition, our
model assumes a constant sticking coefficient, namely
one for all species. The free parameters of our model are
the reaction barriers and the O diffusion efficiency.

Other parameters are the dissociation fraction $\tau$ and the
chemical desorption rate that have been measured previously.
Actually, the chemical desorption could have been neglected in this study, we put it in our model because it increases the quantitative quality of the fitting.
The chemical desorption was already studied in Dulieu et al
2013,\cite{Du13} and in the case of newly formed O$_2$ on silicate
has a value of 40\% $\pm$ 10\%.

The dissociation fraction $\tau$ can be easily calculated by using
this equation:

\begin{equation}
\tau=\frac{(CPS O_2)_{Off}-(CPS O_2)_{On}}{(CPS O_2)_{Off}}*100
\end{equation}
where $(CPS O_2)_{x}$ indicates the counts per second when the
discharge is off or on, with the direct beam passing through the
QMS. Typical values of $\tau$ are between 45-80\%. This
is taken into account through the term  $(1-\tau) \phi$, where
$\phi$ is the normalized flux of O$_2$ molecules when the discharge
is off. Similarly, the term $2\tau \phi$ represents the flux of
oxygen atoms. The rate equations used in our model are:
\begin{align}
\frac{d[O]}{dt} = 2\tau \phi (1-2O\, r_1-O_2\, r_2)-(1-\tau)\phi\, O \,r_2+\\
                 -4k_x \,O O \,r_1 -k_x \,O O_2 \,r_2 \nonumber\\
                 -4k_{td} \,O O \,r_1 -k_{td} \,O O_2 \,r_2 \nonumber\\
\frac{d[O_2]}{dt} = (1-\tau) \phi \,(1-O r_2)- 2\tau \phi\,(O_2 r_2-O r_1)+\\
                   +2k_x \,O O \,r_1 \, (1-\epsilon)-k_x O O_2 r_2\nonumber\\
                     +2k_{td} \,O O \,r_1 \, (1-\epsilon)-k_{td} O O_2 r_2\nonumber\\
\frac{d[O_3]}{dt} = (1-\tau)\phi\, O\, r_2+2\tau \phi \, O_2 \, r_2\\
                   +k_x \,O O_2\, r_2\nonumber\\
                   +k_{td} \,O O_2\, r_2\nonumber
\end{align}
where O, O$_2$, and O$_3$ are the surface densities
(expressed in fraction of ML) of each species, $\phi$ is the flux
(0.003 cm$^{-2}$ s$^{-1}$) of O$_2$, $\epsilon$ is the evaporation
probability due to the chemical desorption, k$_x$ is the diffusion
probability expressed in ML$^{-1}$ s$^{-1}$  (which can be converted into
the usual unit cm$^{2}$ s$^{-1}$ by considering that 1 ML =
10$^{15}$ molecules cm$^{-2}$), and
\begin{align}
r_1 = \nu e^{-\frac{E_{OO}}{k_b T}}\\
r_2 = \nu e^{-\frac{E_{OO_2}}{k_b T}}\\
k_{td} = \nu e^{-\frac{E_d}{k_b T}}
\end{align}
are the O+O and O+O$_2$ reaction probabilities, and thermal
diffusion probability during the heating, respectively.  E$_{OO}$
and E$_{OO_2}$ are the barriers of reactions (3) and (4), E$_{d}$ is
the diffusion barrier (all barriers are expressed in K/k$_b$) and
$\nu$=10$^{12}$ s$^{-1}$ is the trial frequency for attempting a new
event. The diffusion rate k$_x$ (at a fixed T$_s$) includes two
components due to quantum tunneling and thermal motion:\cite{Mi13}
\begin{align}
k_{x} = k_{qt}+k_{tm}
\end{align}
In practice, the diffusion rate during the deposition phase is
governed by a free numerical parameter, whereas the diffusion
coefficient during the heating ramp (k$_{td}$) is described by a
classical thermal hopping mechanism (Arrhenius-type law).
It is possible to use a free parameter during the exposure because
the coverage evolution is known and this represents a strong
constraint, and because the diffusion is supposed to be constant at
constant temperature. However, if the diffusion during the deposition phase followed an
Arrhenius behavior, it would be possible to recognize it a
posteriori. We decided to use the Arrhenius law during the heating
ramp, as most authors did, to compensate for the absence of
constraints on the coverage (O, O$_2$ and O$_3$ populations are not
known at the beginning of desorption) and to describe the evolution
of the diffusion with temperature.

In the rate equations (7), (8), and (9) the ER mechanism is represented by the terms
including the beam flux $\phi$. On the other hand, the LH mechanism
appears in the terms that include the diffusion occurring during the
deposition phase, k$_x$, or during the heating phase, k$_{td}$. By
using this model, we can test, either each at a time or both  at once, the two mechanisms to see how they affect the experimental
observables.

\begin{figure*}[ht]
   \centering
   \includegraphics[width=14cm]{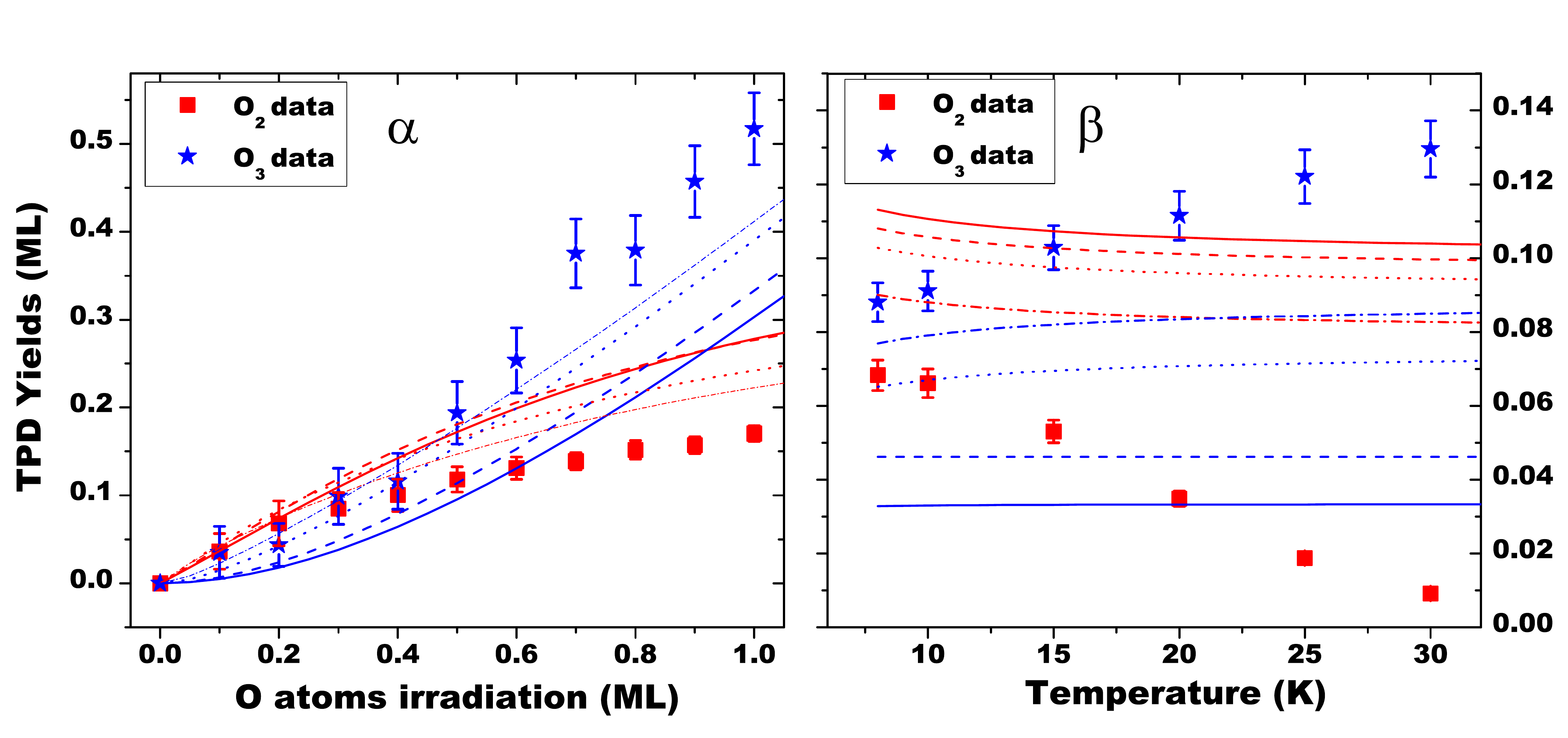}
   \caption{Left panel ($\alpha$): TPD yields vs coverage, comparison between model 1 and experimental data.  Right panel ($\beta$): TPD yields vs surface temperature. Model details: diffusion barrier = 100 K/k$_b$ (solid line), 300 K/k$_b$ (dashed line), 550 K/k$_b$ (dotted line), 900 K/k$_b$ (dashed-dotted line).}
              \label{fig:ModelER}
\end{figure*}

\begin{figure*}[ht]
   \centering
   \includegraphics[width=16.5cm]{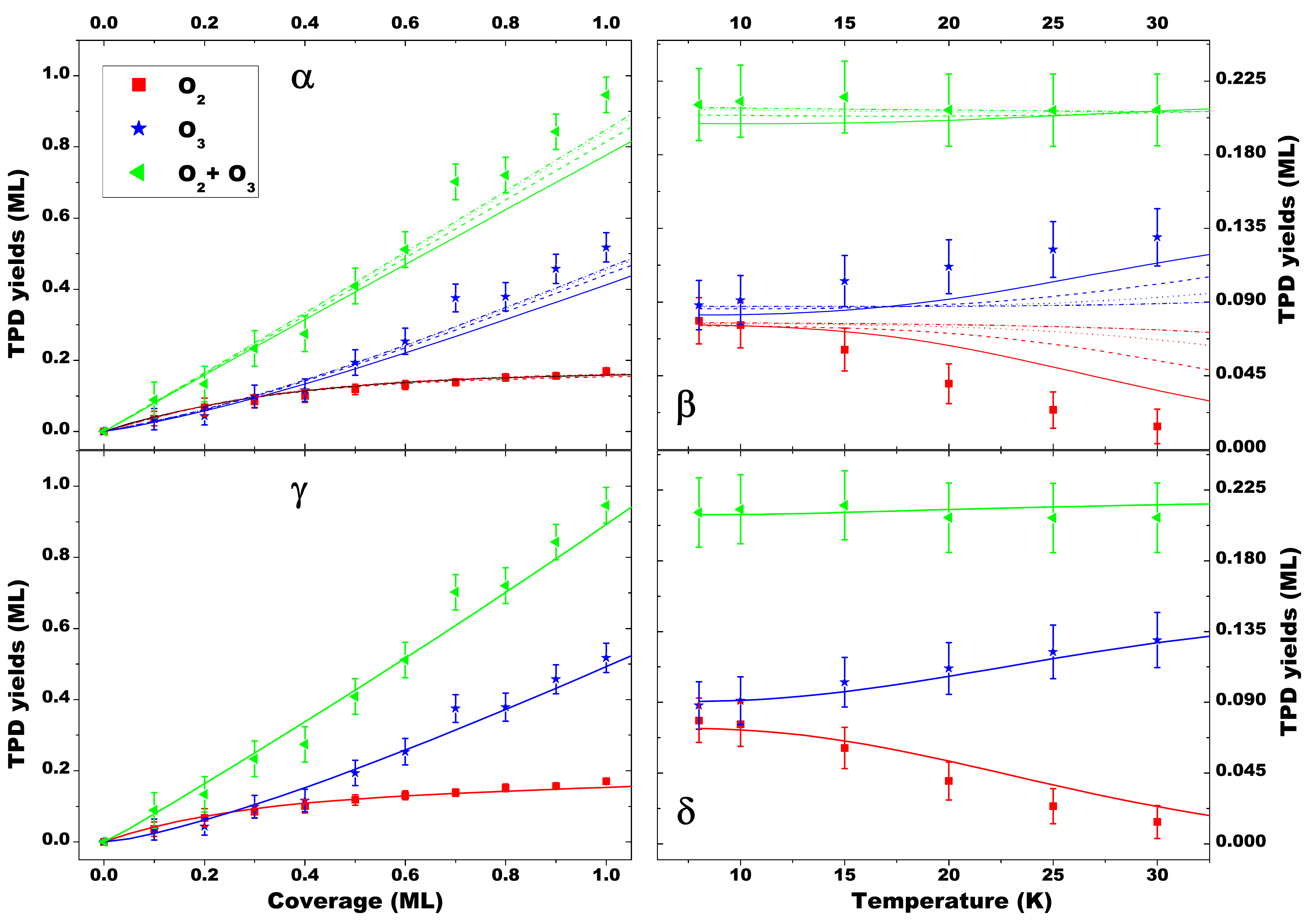}
   \caption{ Comparison between model 2 and experimental data ($\alpha$ and $\beta$ panels). \textit{Case $\alpha$}: TPD yields vs coverage, \textit{Case $\beta$}: TPD yields vs surface temperature (T$_s$). Activation barriers in $\alpha$ and $\beta$: 100 K/k$_b$ (solid line), 300 K/k$_b$ (dashed line), 550 K/k$_b$ (dotted line), 900 K/k$_b$ (dashed-dotted line). Comparison between model 3 and experimental data ($\gamma$ and $\delta$ panels). \textit{Case $\gamma$}: TPD yields vs coverage,, \textit{Case $\delta$}: TPD yields vs surface temperature. Diffusion rates k$_x$(T$_s$) are given in Table 1.}
              \label{fig:ModelLH}
\end{figure*}

\subsection{Model 1: ER and thermal diffusion during TPD with barrieless reactions.}

In the first model, we suppose that reactions occur only through the
ER mechanism (during deposition) or later during the heating phase
(TPD). We assume that there is no diffusion of atoms at low
temperature (k$_x$=0). This is an extreme assumption where the diffusion
cannot occur during the exposure, and especially any tunneling
effect is discarded at the lowest temperatures. We have also assumed
that all reactions are barrierless (except for the O+O$_3$
reaction).\\ In Fig.~\ref{fig:ModelER}$\alpha$ and $\beta$ we
show two results of this model (solid lines) and compare them to the
experimental values. The left panel ($\alpha$) represents the O$_2$
and O$_3$ yield evolution with coverage and the right panel
($\beta$) shows the evolution with temperature. In these models, we
have tested diverse values of the diffusion efficiency during the
TPD, namely, between E$_d$=100 K/k$_b$ (i.e., a very fast
diffusion), and E$_d$=900 K/k$_b$ (i.e., slow diffusion). From model 1, we
conclude that it is not possible to fit our data assuming barrierless reactions and without considering
diffusion during the deposition phase.

 \subsection{Model 2: ER+LH+thermal diffusion with reaction barriers as free parameters.}

Here we include the possibility of diffusion during the exposure
phase and we allow the two reaction barriers (E$_{OO}$ and
E$_{OO_2}$) to vary freely in the range 100 -- 900 K/k$_b$. We then
analyzed the results by applying a minimization method between model
and data for each case. The results are shown in
Fig.~\ref{fig:ModelLH}$\alpha$ and $\beta$. We can see that a
reasonable match was found, although no E$_d$ value satisfies both
coverage and T dependencies at once. In fact, the low diffusion case
(E$_d$=100 K/k$_b$) gives the best fit as far as the temperature evolution
is concerned (Fig.~\ref{fig:ModelLH}$\beta$), but gives the worst fit for the
coverage evolution (Fig.~\ref{fig:ModelLH}$\alpha$), whereas
the high diffusion case (E$_d$=900 K/k$_b$) gives the best fit with
coverage (Fig.~\ref{fig:ModelLH}$\alpha$) and the worst fit
with temperature (Fig.~\ref{fig:ModelLH}$\beta$). As
suggested by the infrared spectra of O$_3$ at 6.5 K, the diffusion of atoms 
during deposition is a key element of the present study.
Hence, it was important to test if the model was able to reproduce
the experimental results obtained during the deposition phase. We
found that experimental values cannot be met if we neglect the
diffusive processes. Moreover, we also demonstrated that we could
have obtained opposite conclusions if we had used only the
temperature, or only the coverage evolutions of the O$_2$ and O$_3$
yields.

\subsection{Model 3: ER+LH+thermal diffusion with barrierless reactions.}

In this model we simulate the same processes seen in the previous
section, but using barrierless reactions (except for the O+O$_3$ reaction).
The model results displayed in Fig.~\ref{fig:ModelLH}$\gamma$
and $\delta$ show a very accurate fit of our data. 
It is important to point out two aspects of these results. First, when
0$<$E$_{OO}$=E$_{OO_2} < 150$ K/ k$_b$, we find several minima in
the $\chi^2$ and we cannot choose a precise value for E$_{OO}$ and
E$_{OO_2}$. Actually, if we assume that there is no barrier to the
reactions, the results are almost identical. From our simulation we
conclude that the activation barriers are so low that they do not slow down reactions (3) and (4) occurring on the surface. 
We can only derive
an upper limit for the two barriers of about 150 K/k$_b$. It is therefore
possible to set the reaction probability equal to one (barrier equal
to zero), and the model remains still fully satisfactory. In Table 1 we
show a series of diffusion parameters we obtained using barrierless
reactions. The diffusion of atoms increases with the temperature,
and follows a T$^n$ law, with n=3 giving the best fit. On the
contrary, using an Arrhenius-type law, this is not possible, or, if we try, the best fit parameters do not have a plausible physical
meaning (i.e., a very low energy and a very low trial frequency).
For this reason, we believe that, on amorphous silicate, as occurs in the
case of water ice,\cite{Mi13} quantum tunneling should be an
important mechanism at low temperature, although we observe a slower diffusion on amorphous silicate than on water ice.

The second important point is that during
the heating phase the diffusion is almost negligible. In fact, not more than a
few \% of the O atoms deposited are still present on the surface in
the very low coverage and temperature regime. For this reason, the
effect of a possible diffusion during the TPD lies within the error
bars of the experimental data. This also validates our assumptions
based on IR spectra recorded at 6 K.

In conclusion, the simplest and most efficient description of our
TPD data is to consider a system that is limited only by the diffusion
during the exposure phase (LH-dominated), and only slightly adjusted
by adding the ER mechanism, especially in regimes of  high coverage ($\geq$ 1 ML).

\begin{table}
\caption{Diffusion coefficients as a function of surface temperature.}              
\label{table:1}      
\centering                                      
\begin{tabular}{c c c c c c c}          
\hline 
\textbf{\backslashbox{k$_{x}$}{T$_{s}$(K)}}   & 8 & 10 & 15 & 20 & 25 & 30   \\    
\hline                                   
\textbf{ ML$^{-1}$ s$^{-1}$} & 0.10 & 0.13 & 0.25 & 0.5 & 1.1 & 2.6   \\      
\hline
\textbf{ cm$^{2}$ s$^{-1}$}$\times$10$^{-16}$ & 1 & 1.3 & 2.5 & 5 & 11  & 26   \\      
\hline                                            
\end{tabular}
\end{table}

\subsection{Evaluation of O+O$_3$ activation barrier}

Here, we want to test whether the reaction O+O$_3$ takes place or not.
In our model, we then added the following terms to the right side of
Eq. (7), (8), and (9), respectively:

\begin{align*}
-2\mu \phi O_3 r_3 -k_{td} O O_3 r_3-k_x O O_3 r_3  \\
2 (\mu \phi O_3 r_3 + k_{td} O O_3 r_3 + k_x O O_3 r_3)\\
-2\mu \phi O_3 r_3 -k_{td} O O_3 r_3 -k_x O O_3 r_3
\end{align*}
where
\begin{align}
r_3 = \nu e^{-\frac{E_{OO_3}}{k_b T}}
\end{align}

is the O+O$_3$ reaction probability and E$_{OO_3}$ is the barrier to
reaction (5). By varying E$_{OO_3}$, we noticed that even
for values bigger than $\sim$ 2000 K/k$_b$, the amount of O$_2$ and
O$_3$ remained unaltered; while for values of E$_{OO_3} < 2000 K/k_b$
the $\chi^2$ started to increase. To give an estimate of a lower
limit of the barrier, we took the error bars as the borders limiting a 
the maximum allowed amount of O$_2$ yield (the
smaller E$_{OO_3}$ the more O$_2$ is produced). We found a reaction
barrier of 2300 K/k$_b$. This means that the barrier is very likely
to be greater than this value. In addition, this value is consistent
with the data available in gas phase, where barriers were found to
be greater than 1950 K/k$_b$.\cite{Wi83} This convincingly shows
that the O+O$_3$ reaction is slow enough that it can be neglected
in our model.

\subsection{Hot Atom mechanism}

\begin{figure*}
   \centering
   \includegraphics[width=16.5cm]{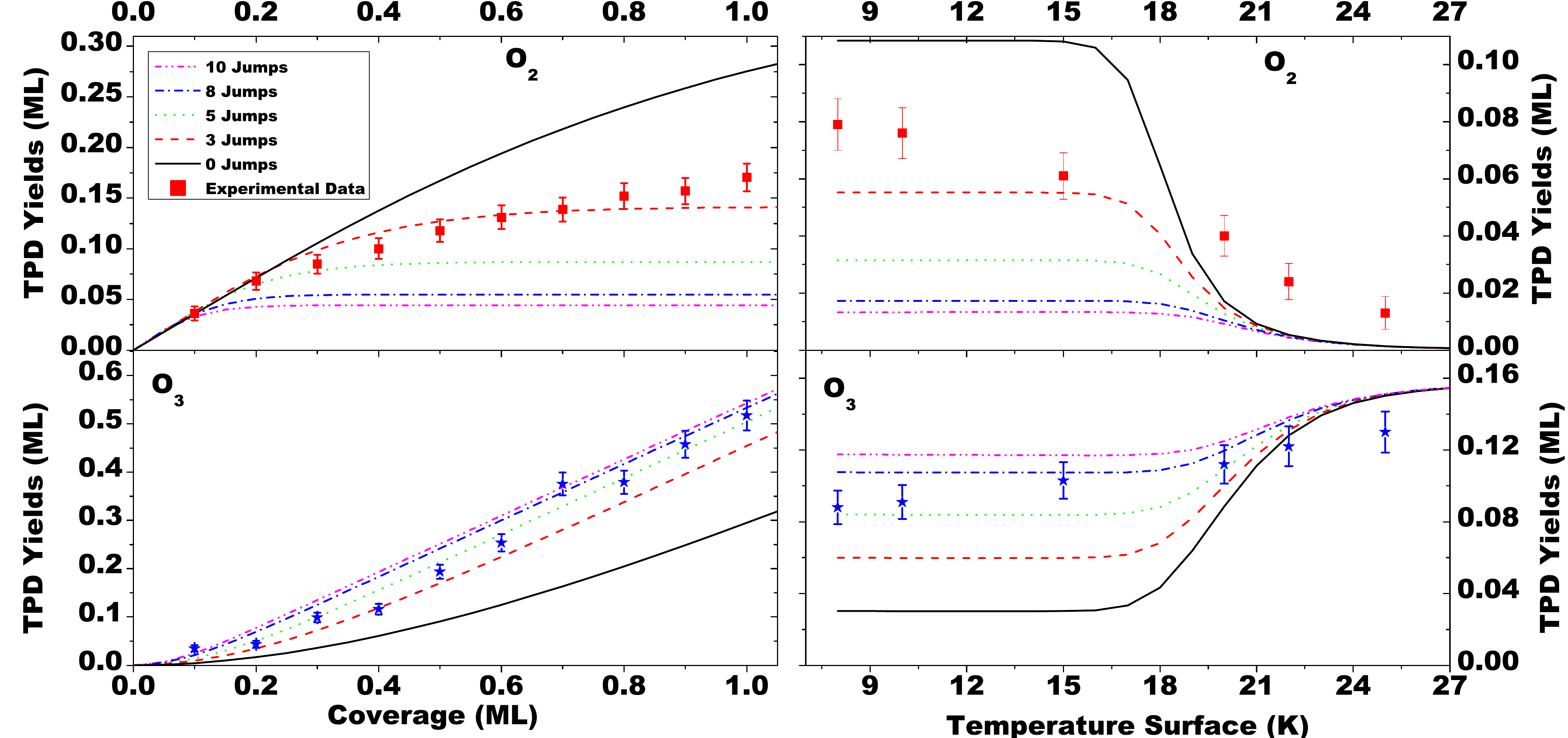}
   \caption{Comparison between model 2 + Hot Atom mechanism and experimental data. \textit{Left panel}: O$_2$ (top) and O$_3$ (bottom) TPD yields vs coverage. \textit{Right panel}: O$_2$ (top) and O$_3$ (bottom) TPD yields vs surface temperature. Model results (lines) are generated by considering 5 cases: HA does not take place (0
jumps, solid line) or HA occurs and the atoms are able to scan 3, 5,
8 or 10 adsorption sites (dash, dot, dash-dot and dot-dash-dot
curves, respectively).}
              \label{fig:HotAtom}
\end{figure*}

Generally, only ER and LH mechanisms are considered when the
formation of molecules via surface chemistry is concerned. However,
a molecule arriving at the surface may not be chemisorbed (or
physisorbed) upon the first impact due to the inefficient energy
transfer between the impinging particle and the surface. Before the
complete dissipation of its incident energy, the adsorbed particle
is not in thermal equilibrium with the surface. Hence, impinging
particles could be able to hop on the surface and react with already
adsorbed molecules lying several angstroms away from the impact
site. In the literature, this process is called ``Hot Atom'' (HA) or
Harris-Kasemo\cite{Ha81} mechanism. To date, HA has been studied
mainly from a theoretical point of view (Martinazzo et al
2004,\cite{Ma04} and Molinari \& Tomellini 2002,\cite{Mo02} and ref.
therein), although some experimental studies exist (Wei \& Haller 1996,\cite{We96} and
Dinger et al 2001,\cite{Di01} and ref. therein). Previous works
considered metallic surfaces only, and atoms with an energy greater
than 0.5 eV or light atoms (H or D); under these conditions, the
energy transfer between the particles and the surface is slow, so
there is a high probability that the HA mechanism occurs. In our
experiments, we worked under very different conditions. We performed the experiments on
non-metallic surfaces (silicate, graphite, and water ice), atoms had an
energy $<$ 0.01 eV and were heavy particles (O atoms,
mass(O)/mass(H)=16). These
considerations lead us to assume that the HA mechanism should not be
important under our experimental conditions, especially at low
coverages. Another problem -- still unsolved to date -- is
the surface temperature dependence of the HA mechanism. Some
experimental and theoretical works (Quintas-S\'anchez et al
2013,\cite{Qu13} and ref. therein) show a temperature dependence of
HA, but the range of temperature used is very broad (more than 300
K). In our case, the range of the surface temperatures is small
($<$ 25 K) and it is reasonable to assume that the energy transfer between an
adsorbed particle and the surface is quite constant within this
range of temperatures. We tried to include the HA mechanism in our
model taking into account all the points discussed above. It turns out
that the HA mechanism does not exhibit a surface temperature dependence under our experimental conditions, and we
may consider it as an ``enhanced ER mechanism'': atoms coming from
the gas phase are likely to scan more than one adsorption site and
thus have a higher probability to react. In
Fig.~\ref{fig:HotAtom}, we show the results of the model for
O$_2$ and O$_3$ yields vs coverage (left panel) and O$_2$ and O$_3$
yields vs surface temperature (right panel). The curves shown in the
figure were generated by considering 5 cases: HA does not take place
(0 jumps, solid line) or HA takes place and the atoms are able to
scan 3, 5, 8 or 10 sites. The traces displayed in the left panel
(3-5 jumps) seem
to fit the experimental data 
although this is not the case for the curves in the right
panel, where the temperature independence of ER and HA prevents the
model from converging to a good fit for O$_2$ and O$_3$. In this
case, the temperature dependence is contained only in the Arrhenius
term used to simulate the TPD. Regardless of the diffusion barrier
we may use (in the case shown in Fig.~\ref{fig:HotAtom}, E$_d$ is
500 K/k$_b$), we are not able to reproduce the plateau behavior
of the experimental values. In fact, the exponential law
induces a sudden change (shifted towards low or high temperature
depending on the value of E$_d$) in the yields O$_2$ and O$_3$.

The lack of theoretical and experimental studies about HA under the
same conditions used in the present work (a non-metallic surface, low
energy and heavy atoms), together with the results of our model,
suggests that HA cannot explain the experimental results. This corroborates the hypothesis that the diffusion mechanism (LH) is dominant at low temperatures and in low coverage regimes.

\section{Conclusions and astrophysical implications}

In this paper we have shown that O$_3$ can be formed very
efficiently via an atomic oxygen beam sent on silicate held at low
temperatures (6-30 K). The reactions leading to ozone formation
studied in this paper (O+O and O+O$_2$) appear to be barrierless; we
have estimated an upper limit for the activation energies of reactions of 150 K/k$_b$. Conversely,
 the reaction O+O$_3$ has a high activation barrier --
lower limit $\sim$ 2300 K/k$_b$ -- and it is not an efficient
pathway for ozone destruction under our experimental conditions. In
addition, the formation of ozone is favored by a very fast
diffusion of oxygen atoms at low temperatures. The diffusive process
of O atoms is likely to occur via quantum tunneling, as claimed in
Minissale et al 2013,\cite{Mi13} while the Hot Atom mechanism effects proved to be negligible.

From an astrophysical point of view, since the gas phase abundances
of O and O$_2$ are elusive,\cite{Je09, Wh10} it is difficult to put
the O$_3$ formation in a simple interstellar contest. New and more
detailed observational data are necessary to know the solid phase
abundances of ozone. 
Anyway, we can fairly assume that in dense clouds, particularly UV-protected interstellar environments (A$_v$ $>$ 3), if comparable budgets of O atoms and H atoms are present,\cite{Ca02} an O-addition chemistry competes with H additions and O$_3$ could be formed.
The presence of ozone in the interstellar ices
would confirm that the O$_3$+H pathway is the most important route
leading to water in some interstellar environments.
Not only ozone is important because it
represents an efficient way to produce water, but also because it
can be a reservoir of oxygen atoms; due to its low binding energy
(1.1 eV), ozone can be easily processed by cosmic-rays and by the
mean Galactic UV field,\cite{Ti82, St91} producing O atoms that, for
example, could react on the surface of dust grains with CO and
produce CO$_2$.\cite{Ma06,Ra11} Even without energetic events,
CO+OH\cite{Ob10, No11} and CO+O\cite{Mi13b} reactions are believed
to be sources of CO$_2$. The observed concomitance of CO$_2$ and
H$_2$O in the ices\cite{Wh10} can be more easily understood by
assuming that this chemistry at the surface of dust grains is
driven by the presence or absence of O$_3$. Ozone can be either an
OH provider via its hydrogenation, or an O consumer upon its own
formation (O $+$ O$_2$). Therefore, oxygen diffusion and reactivity
on cold surfaces are also key factors to understanding the CO$_2$ and H$_2$O
formation rates.

\begin{acknowledgments}
\textit{FD, EC, and MM acknowledge the support of
the national PCMI programme founded by CNRS, the Conseil Regional
dIle de France through SESAME programmes (contract I-07597R). The
authors thank S. Baouche for his valuable technical assistance. MM
acknowledge LASSIE, a European FP7 ITN Communitys Seventh Framework
Programme under Grant Agreement No. 238258. MM also thanks Prof.
Tomellini from the ``Universit\`a di Roma Tor Vergata'' for fruitful discussions.}\end{acknowledgments}

\end{document}